\documentstyle[12pt, epsf]{article}
\def\by#1#2{{\displaystyle {#1}\over \displaystyle {#2}}}
\def\d{{\rm d}}

\def\gev2{\hbox{GeV}^2}
\def\<{\langle}
\def\>{\rangle}
\begin{document}

\begin{flushright}
DO-TH-95/21 
\end{flushright}

\begin{center}
\Large \bf
Nuclear Modification of Double Spin Asymmetries \\ [1.2cm]
\normalsize D. Indumathi, \\ [0.3cm]
Instit\"ut f\"ur Physik T-IV,
Universit\"at Dortmund, D-44221, Germany\footnote{e-mail:
indu@hal1.physik.uni-dortmund.de}
\end{center}

\vspace{1.0cm}

\begin{abstract} 
{\noindent We compute nuclear spin dependent structure
functions using a dynamical model for bound nucleon densities and
hence calculate nuclear modifications to asymmetries observed in recent
doubly polarised deep inelastic scattering experiments. We conclude
that while the individual densities are changed substantially by nuclear
effects, the asymmetries themselves are largely insensitive to these
changes.}

\end{abstract}

Recently a model was proposed \cite{IZ} to explain the observed
differences between free nucleon and bound nucleon structure functions
in deep inelastic lepton nucleon scattering (DIS). This model used a
dynamical approach, involving modifying a free nucleon input density
distribution at a low input scale, $Q^2 = \mu^2 = 0.23$ GeV${}^2$, due
to nuclear effects, and then evolving the resultant modified bound
nucleon densities to the required $Q^2$ scale of the experiment. The
model gave satisfactory agreement with available data in a fairly broad
$Q^2$ range, from 0.5--30 GeV${}^2$. 

It is interesting to ask how this model can be extended to a study of
{\it spin dependent} bound-nucleon densities. The question is not merely
academic as, in fact, data on the spin dependent deuteron and neutron
structure functions have been obtained \cite{EMC} from deuteron and
${}^3$He targets. Nuclear effects in deuteron are known to be small
(though measureable), since the deuteron is a loosely bound nucleus.
There have been a number of papers \cite{spind} dealing with nuclear
modifications of spin asymmetries and structure functions in the case of
the deuteron\footnote{Depending on the model, corrections due to nuclear
effects in deuterium can be as large as 10\%.}. We therefore confine our
attention to possible nuclear effects on the double spin asymmetry
measurements made with helium nuclei. In this case, it was pointed out
by Woloshyn \cite{Wolo} that the protonic contribution to the asymmetry
is negligible so that the ${}^3$He double spin asymmetry is sensitive
to the spin dependent neutron structure function, $g_1^n(x, Q^2)$.
However, there may be additional modifications due to the presence of
the nuclear medium, which we propose to study here. These are especially
of importance for checking the validity of the Bjorken Sum rule. Our
main conclusion is that the individual (spin independent as well as
spin dependent) structure functions undergo substantial modifications
due to nuclear effects; however, their {\it ratio}---the asymmetry---which
is the measured quantity, is largely free from these and so gives hope that
the neutron structure function may be unambiguously determined from such
a measurement. 

The quantity of interest is the double spin asymmetry, 
$$
A_1^A (x, Q^2) = \by{g_1^A(x, Q^2)}{F_1^A(x, Q^2)}~,
\eqno(1)
$$
where $g_1$ and $F_1$ are the spin dependent and spin independent
structure functions corresponding to the nucleus $A$. We are therefore
interested in studying possible deviations of the measured asymmetry,
$A_1^{\rm He}$, from the required neutron asymmetry, $A_1^n$, due to
nuclear effects. 

As a starting point we note that the corresponding study of unpolarised
bound nucleon densities used the GRV \cite{GRVunpol} density
parametrisation as an input. We shall therefore use the GRVs \cite{GRVpol}
spin dependent densities as an input in the corresponding polarised
problem. This is essential if we are to retain the definition,
$$
\begin{array}{rcl}
q_f(x) & = & q_f^+(x) + q_f^-(x) ~; \\
\tilde{q}_f(x) & = & q_f^+(x) - q_f^-(x) ~, 
\end{array}
\eqno(2)
$$
where $q_f^+$ and $q_f^-$ are the positive and negative helicity
densities of $f$--flavour quarks (and similarly for gluons) in either
free or bound nucleons. 

\section{The Spin Independent Nuclear Densities}
We now quickly review the model before we apply it to the polarised
case. This is also useful as the polarised case essentially follows
along the lines of the unpolarised problem. The free nucleon input
densities are modified by both nuclear swelling and binding effects.
Nucleon swelling causes not only a depletion of {\it all} parton
densities at large- and small-$x$, but also an enhancement at
intermediate-$x$ \cite{Zhu}. The relative increase in the nucleon's
radius is $\delta_A$, where $(R_N+\Delta R(A))/R_N = 1 + \delta_A$, and
is given by,
$$
\delta_A=[1-P_s(A)]\delta_{\rm vol}+P_s(A)\delta_{\rm vol}/2~.
$$
The second term corrects for surface effects in the usual manner. 
Here $P_s(A)$ is the probability of finding a nucleon on the nuclear
surface while $\delta_{\rm vol}$ parametrises the swelling of the
nucleon in the interior of a heavy nucleus and is the only free
parameter in the calculation. It was fixed to be $\delta_{\rm vol}
= 0.15$ in the unpolarised calculation \cite{IZ}. 

The distortions of the density distributions due to swelling, being
purely geometrical, conserve the total parton number and momentum of
each parton species (i.e., the first and second moments of the
distributions are unchanged). Furthermore, the third moments are
modified in a well-determined way. Specifically, the first three moments
of the parton distributions in a free ($q_N$) and bound ($q_A$) nucleon
at $\mu^2$ are related by
$$
\begin{array}{rcl}
\langle q_A(\mu^2)\rangle_1  & = & \langle q_N(\mu^2)\rangle_1~, \\
\langle q_A(\mu^2)\rangle_2  & = & \langle q_N(\mu^2)\rangle_2~, \\
{\displaystyle (\langle q_N(\mu^2)\rangle_3-\langle q_N(\mu^2)
     \rangle_2^2)^{1/2} \over \displaystyle (\langle q_A(\mu^2)
     \rangle_3-\langle q_A(\mu^2)\rangle_2^2)^{1/2}} & = & 1+\delta_A~.
\end{array}
\eqno(3)
$$
The first two equations imply number and momentum conservation of 
partons and the last incorporates the swelling effect \cite{Zhu}.

These are then used as constraint equations to determine the
bound-nucleon densities in terms of the known free-nucleon ones. We
use the Gl\"uck, Reya and Vogt (GRV) parametrisation \cite{GRVunpol}
of the free-nucleon distributions at $Q^2 = \mu^2 = 0.23 \,{\rm GeV}^2$.
We find these most appropriate for our purpose as each of their input
densities is integrable (there are finite number of partons at $\mu^2$). 

It is now possible to determine the bound nucleon densities $q_A$ (for
$q$ = valence quarks, $u_V$, $d_V$, sea quarks, $S$, and gluons, $g$)
in terms of the free densities, $q_N$. We parametrise the free as well
as bound nucleon distributions in the form,
$$
q(x) = N x^\alpha (1-x)^\beta P(x),
\eqno(4)
$$
where $P(x)$ is a polynomial. We take $P_{A, q}(x) = P_{N, q}(x)$, for
simplicity. Then the changes in the three main parameters, $N$, $\alpha$,
and $\beta$, due to swelling, and hence the bound-nucleon densities, are
immediately determined by the constraints in eq (3). This fixes the input
bound-nucleon densities at $Q^2 = \mu^2$.

We now discuss the binding effect. The attractive potential describing the 
nuclear force arises from the exchange of mesons. Hence
the energy required for binding is taken away solely from the mesonic
component of the nucleon, and not from its other components. At the
starting scale, $Q^2=\mu^2$, we identify these mesons to be just the sea
quarks in the bound nucleon. Hence, the momentum fraction carried by the
sea quarks in a nucleon bound in a nucleus at $Q^2=\mu^2$ will be reduced.
The extent of reduction is determined by the binding energy per nucleon
\cite{IZ}, which is given by the well-known Weiz\"acker mass formula. 

These nuclear effects completely determine the spin independent input
bound-nucleon densities. These are then evolved using the usual
Altarelli Parisi evolution equations to obtain the densities at a
required value of $Q^2$. 

At the time of interaction, there is a further depletion of the sea
densities, which occurs whenever there is nucleon-nucleon interaction,
caused by parton--nucleon overlap. When a parton having a momentum
fraction, $x$, of the parent nucleon momentum, $P_N$, is struck, it is
off-shell and localised to a distance, $\Delta Z\sim 1/(2xP_N)$ (in
the Breit frame). For sufficiently small $x$, $\Delta Z$ becomes large
and can exceed the average 2--nucleon separation\footnote{Although the
spatial extent of a single coloured parton cannot exceed the range of
QCD confinement, the struck parton can combine with a wee parton 
and form a colourless scalar with vacuum quantum numbers which can then
escape from the nucleon.}. 

The struck parton must return to the parent nucleon within the
interaction time, as required by the uncertainty principle. However, 
while it extends outside the parent nucleon, it can interact with other
nucleons in the nucleus. Such an interaction between two nucleons caused
by parton--nucleon overlap results in loss of energy of the parent
nucleon, mimicking exactly the effect of binding. Hence we call this the
second binding effect and assume its strength to be the same as that
due to the usual binding. This immediately fixes the loss in sea quarks
(valence quarks are not depleted due to the requirement of quantum
number conservation) due to this effect to be 
$$
S'_A(x, Q^2) = K'(A) S_A(x, Q^2)~,
\eqno(5a)
$$
where the depletion factor is,
$$
\begin{array}{rcll}
K'(A) & = & 1, & \hbox{when}~x > x_0; \\
& = & 1-2\beta(x_0x^{-1}-1), & \hbox{when}~x_A < x < x_0; \\
& = & 1-2\beta(x_0x_A^{-1}-1), & \hbox{when}~x < x_A,
\end{array}
\eqno(5b)
$$
where $\beta$ is the same as in usual binding, viz., 
$$
\beta = \frac{U(\mu^2)}{M_N\langle S_N(\mu^2)\rangle_2} = 0.037/2~,
\eqno(6)
$$
$U(\mu^2)$ being the binding energy between each pair of nucleons, which
is known. The limiting values, $x_0 = 1/(2M_N d_N)$ (where $d_N$
is the average correlation distance between two neighbouring nucleons
in the lab frame), and $x_A=1/(4\overline{R}_A M_N)$ (where 
$2 \overline{R}_A \simeq 1.4 R_A$ is the average thickness of the
nucleus), determine the starting and saturation values respectively of
this shadowing effect; the latter occurs when the struck quark wave
function completely overlaps the nucleus in the $z$-direction. In
general, a parton with a momentum fraction, $x$, $x_A \le x \le x_0$, can
overlap $(n-1)$ other nucleons, where $n=1/(2 M_N d_N x) = x_0/x$.
Due to the applicability of the superposition principle to the scalar
field interaction with various nucleons, the loss of energy due to
interaction with each of the nucleons over which the struck quark
wave function extends, is equal and additive, and thus explains the
depletion factor in eq (5). Since this effect acts on the
intermediate state of the probe--target interaction, it does not
participate in the QCD evolution of the initial state. 

Nuclear modification due to binding and swelling at the input scale $Q^2
= \mu^2$, and parton-nucleon overlap due to the second binding effect at
the $Q^2$ scale of the scattering together determine the structure
function, $F_1^A(x, Q^2)$, of a nucleon bound in a nucleus $A$. The
model gives good agreement with available data \cite{IZ}. 

We now proceeed to an analysis of the corresponding spin dependent
densities. 

\section{The Spin Dependent Nuclear Densities}
The same nuclear effects of binding and swelling affect the spin
dependent densities also. This is because they influence the positive
and negative helicity densities, out of which the spin independent and
spin dependent densities are composed (see eq (2)). The entire swelling
effect can now be rephrased as the effect of swelling on individual
{\it helicity} densities, so that equations analogous to (3) are valid
for the spin dependent densities, $\tilde{q}(x)$, as well. This can be
seen as follows: Swelling simply rearranges the parton distributions in
the bound nucleon; there is no change in the number of each parton
species. In particular, each helicity type is also conserved, i.e., 
$$
\int q_A^+ (x, \mu^2) \d x= \int q_N^+ (x, \mu^2)\d x ~, ~~~~~~~~
\int q_A^- (x, \mu^2) \d x= \int q_N^- (x, \mu^2)\d x ~. 
$$
Hence, their sum and difference is also conserved. The former is
contained in the first equation of the equation set (3); the latter
implies, for the polarised combination,
$$
\langle \widetilde{q_A} (\mu^2) \rangle_1 = 
\langle \widetilde{q_N} (\mu^2) \rangle_1~.
\eqno(7a)
$$
Note that $\langle q(\mu^2)\rangle_n = \langle q^+(\mu^2)\rangle_n
+ \langle q^-(\mu^2)\rangle_n$ for every moment, $n$, for both the free
and bound nucleon, and similarly for the spin dependent density as well.
Similarly, since the momentum carried by each helicity density is
unchanged, momentum conservation between the free and bound nucleon also
holds for the sum and difference of the helicity densities. The
corresponding equation for the sum is the second equation in (3); the
equation for the helicity difference is
$$
\langle \widetilde{q_A} (\mu^2) \rangle_2 = 
\langle \widetilde{q_N} (\mu^2) \rangle_2~.
\eqno(7b)
$$
The extension of the third of the equations in (3) to the spin dependent
case is not as straightforward. Every helicity density, $q_f^h(x)$, 
$(h = +, -)$, spreads out over a larger size, or, equivalently, gets
pinched in momentum space, according to Heisenberg's uncertainty
relation, $\Delta p \Delta x = 1$. Applying this to each helicity type,
for each flavour, we have, 
$$
{\displaystyle (\langle q_N^+(\mu^2)\rangle_3-\langle q_N^+(\mu^2)
     \rangle_2^2)^{1/2} \over \displaystyle (\langle q_A^+(\mu^2)
     \rangle_3-\langle q_A^+(\mu^2)\rangle_2^2)^{1/2}} = 1+\delta_A~;
     ~~~~~
{\displaystyle (\langle q_N^-(\mu^2)\rangle_3-\langle q_N^-(\mu^2)
     \rangle_2^2)^{1/2} \over \displaystyle (\langle q_A^-(\mu^2)
     \rangle_3-\langle q_A^-(\mu^2)\rangle_2^2)^{1/2}} = 1+\delta_A~.
\eqno(8)
$$
However, for later convenience, we prefer to use analogous expressions
for the sum and difference, $q_f$ and $\tilde q_f$, rather than for the
individual helicity densities. Hence, the third of the constraints
arising from swelling, i.e., the third of eq (3) and its spin dependent
counterpart read,
$$
{\displaystyle (\langle q_N(\mu^2)\rangle_3-\langle q_N(\mu^2)
     \rangle_2^2)^{1/2} \over \displaystyle (\langle q_A(\mu^2)
     \rangle_3-\langle q_A(\mu^2)\rangle_2^2)^{1/2}} = 1+\delta_A~;
     ~~~~~
{\displaystyle (\langle \widetilde{q_N}(\mu^2)\rangle_3-\langle 
\widetilde{q_N}(\mu^2) \rangle_2^2)^{1/2} \over \displaystyle (\langle 
\widetilde{q_A}(\mu^2) \rangle_3-\langle \widetilde{q_A}(\mu^2)
\rangle_2^2)^{1/2}} = 1+\delta_A~.
\eqno(7c)
$$
The error involved between the exact expressions, eq (8), and their
approximations, eq (7c), is a term proportional to $(1 - (1 + \delta_A)^2)$
and is of order $\delta_A$. This term mixes spin dependent and spin
independent moments; however, since $\delta_A$ is small (about 10\%),
these errors are small, and can be ignored. We are therefore justified in
using eq (7c) rather than eq (8) to constrain the second moments of
the parton densities. The three sets of equations, (7a--c), thus provide
the three sets of constraint equations, analogous to the set (3), with
which we can fix the input bound nucleon spin dependent densities. 

The modified input densities are thus determined, given a set of valid
input free nucleon distributions, which we take to be the Gl\"uck,
Reya, and Vogelsang 'standard' set (GRVs) \cite{GRVpol}. These densities
can also be parametrised in a form similar to eq (4); in fact, every
spin dependent density is a factor of the form of the RHS of eq (4)
times the corresponding unpolarised density. Hence there are again three
constraint equations which serve to fix the three main parameters,
$\alpha$, $\beta$, and $N$ for the corresponding bound nucleon spin
dependent densities.

Binding causes loss of energy in the sea: this is due to loss of mesons
from the nucleon. Since these mesons are spin-0 bosons, it is clear
that no spin is lost from the sea due to binding (equal numbers of
positive and negative helicity partners are lost). Hence we see that
binding changes the sum, but not the difference of the helicity
densities\footnote{It is possible that $\rho$, etc., mesons also
participate in this interaction, leading to a change in the polarised
sea densities, but this component is small and we neglect it.}.

We thus obtain the input polarised densities analogous to the
unpolarised ones. These are then evolved to the scale of interest.

At the time of interaction, the second binding effect applies to struck
partons with momentum fraction $x \le x_0$, as in the unpolarised case. 
The mechanism for this depletion is independent of the helicity of the
quark, and so this effect is identical in both the spin independent as
well as spin dependent cases. Hence, the spin dependent structure
function, $g_1^A(x, Q^2)$ can now be computed by evolving the modified
input spin dependent densities to the required value of $Q^2$, and
including the second binding effect. 

Finally, we display the equivalent neutron bound-nucleon structure
functions at an arbitrary scale, $Q^2 > \mu^2$ (with $R =
\sigma_L/\sigma_T = 0$) :
$$
\begin{array}{rcl}
$$
F_1^{n/A}(x,Q^2) & = & \by{1}{18} \left[u_v^A(x,Q^2) + 4\, d_v^A(x,Q^2)
	               + K'(A) S_A(x,Q^2) \right]~, \\
g_1^{n/A}(x,Q^2) & = & \by{1}{18}  \left[\widetilde{u_v}^A(x,Q^2) + 
		        4\, \widetilde{d_v}^A(x,Q^2) 
                        + K'(A) \widetilde{S}_A(x,Q^2) \right]~.
\end{array}
\eqno(9)
$$
$q^A(x,\mu^2)$ incorporates the effect of swelling on every input parton
density, $q^N(x,\mu^2)$, as well as that of binding for the unpolarised
densities, and the corresponding $Q^2$-dependent quantities that appear
here are these input densities, evolved suitably to the required scale.
$K'(A)$ incorporates the second binding effect, at $Q^2$, as discussed
above. The experimentally measured asymmetry, and quantity of interest,
are the ratios, at the scale $Q^2$, for the neutron bound in the helium
nucleus and for a free neutron:
$$
{\cal{A}}^{\rm meas} = \by{g_1^{\rm n/He}}{F_1^{\rm n/He}}~, \hspace{3cm} 
{\cal{A}}^{\rm reqd} = \by{g_1^{\rm n}}{F_1^{\rm n}}~,
\eqno(10)
$$
and can thus be computed. (We use the free and bound nucleon {\it
unpolarised} structure functions from \cite{IZ}). Note that the input
spin dependent densities (which are taken from \cite{GRVpol}) were
actually fitted to both the free proton as well as deuteron and
${}^3$He spin dependent data; however, we use them here as
the free nucleon parametrisations (which is
permissible especially in view of the large error bars on presently
available data). Furthermore, the smearing effect of Fermi motion (at
large $x$) is neglected in this work for simplicity. Hence our results
are not valid at large $x$. 
	
In fig.\ 1 we give the results of our computations for the measured
(bound nucleon) and required (free nucleon) spin dependent structure
function, $g_1^n$ for typical values of $Q^2$, $Q^2 = 1, 4$ GeV${}^2$.
We see that the deviations of the bound neutron structure function can
be as large as 10--15\% at small $x$ and about 6\% at intermediate $x$
values. The data points plotted on this graph correspond to the values
extracted at $Q^2 = 4$ GeV${}^2$ from a measurement of the asymmetry by
the E142 Collaboration \cite{spind} (with $R = 0$) and indicate the size
of the error bars in currently available data. In fig.\ 2, we plot the
asymmetries at $Q^2 = 4$ GeV${}^2$. The data points here correspond
exactly to the E142 data and therefore go over a range of $Q^2$ with a
mean of about 2 GeV${}^2$; however, the asymmetry is not very sensitive
to $Q^2$ in the $x$ range of the available data. Notice that in this
case, nuclear effects cause not more than 5\% deviation in
the asymmetry at both small and intermediate values of $x$. The deviation
is slightly larger at larger $x$, $x > 0.4$, but this is due to the fact
that the neutron spin dependent structure function changes sign near this
value, and hence this deviation cannot be considered to be significant. 

In short, we see that nuclear effects, though significant, equally
affect both the spin dependent as well as the spin independent structure
functions in such a way that the measured asymmetries are to a great
extent independent of them. Since it is the asymmetry rather than the
structure function which is measured in a polarised experiment, much
smaller errors on data are required before these small deviations due to
nuclear effects become observable in such experiments. On the other
hand, as already stated, this seems to make possible clean and
unambiguous extraction of the relevant {\it free nucleon} structure
functions from a measurement of double spin asymmetries with such
light nuclear targets.

{\bf Acknowledgements}: I thank M. Gl\"uck for suggesting the idea that a
computation of nuclear effects on spin structure function data may be
interesting; I thank E. Reya for a critical reading of the manuscript, 
and M. Stratmann for providing the relevant {\sc fortran} programs.

\noindent {\Large \bf Figure Captions}

\vspace{0.8cm}

\noindent Fig. 1 The free and bound nucleon spin dependent structure
function for $Q^2 = 1, 4 $ GeV${}^2$ as a function of $x$ are shown as
solid and dashed lines respectively. The structure function data are
extracted at $Q^2 = 4$ GeV${}^2$ from the asymmetries measured by the 
E142 collaboration. 

\vspace{0.3cm}

\noindent Fig. 2 The bound and free nucleon asymmetries for $Q^2 = 4 $
GeV${}^2$ as a function of $x$ are shown as solid and dashed lines
respectively. The data are from the E142 collaboration. 

\vspace{0.3cm}

\begin{center}
\leavevmode
\epsfxsize=5.0in 
\epsfbox{g1n.prn}

\vspace{0.5cm}

Fig. 1

\end{center}

\vspace{0.3cm}

\begin{center}
\leavevmode
\epsfxsize=5.0in 
\epsfbox{a1n.prn}

\vspace{0.5cm}

Fig. 2

\end{center}

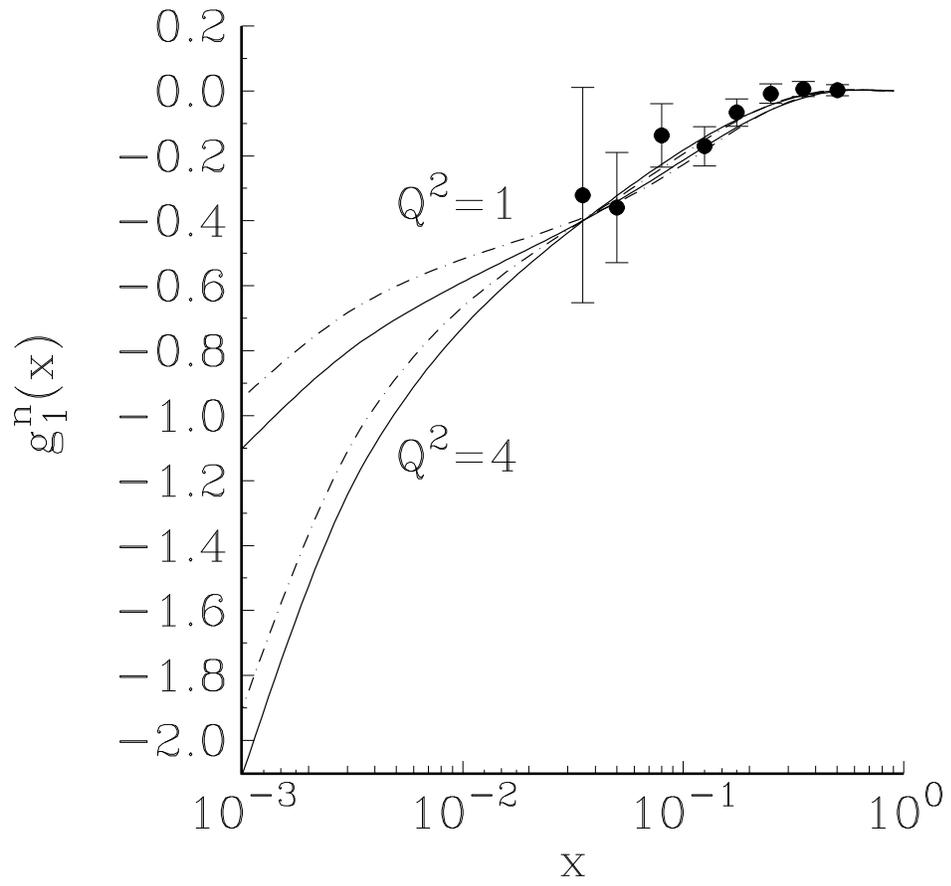
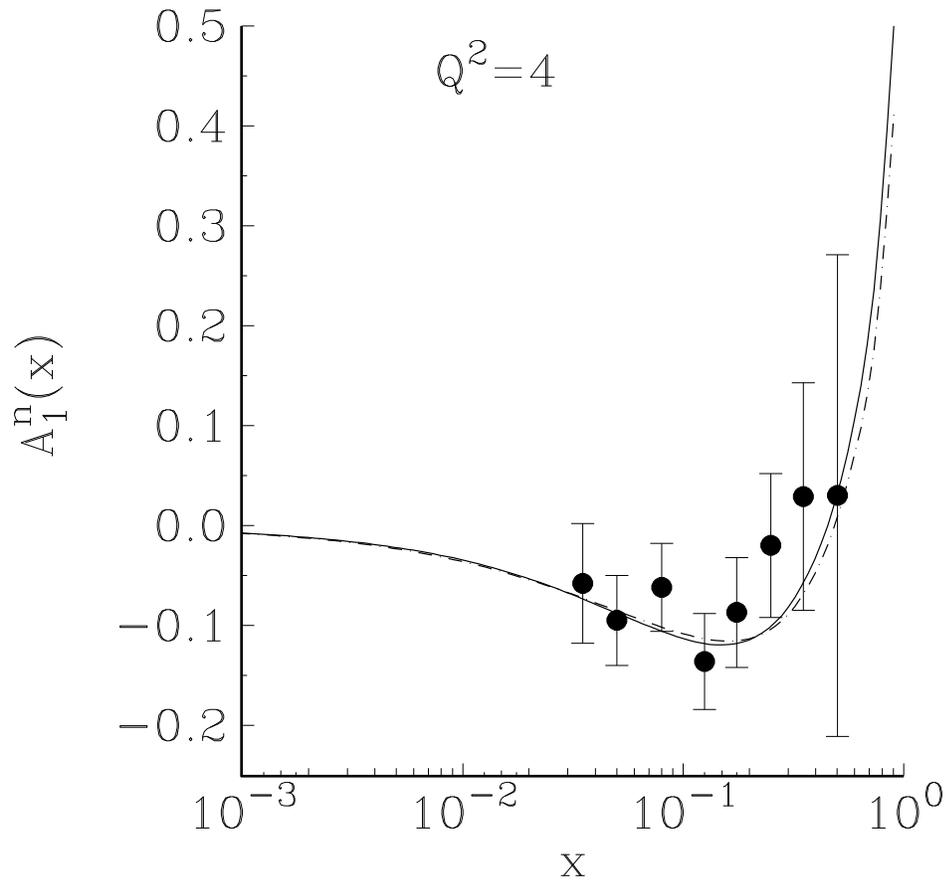
\end{document}